\documentclass[italian,english]{article}
\usepackage[latin9]{inputenc}
\usepackage{color}
\usepackage{graphicx}
\usepackage{amssymb}

\newcommand{\lyxaddress}[1]{
\par {\raggedright #1
\vspace{1.4em}
\noindent\par}
}

\usepackage{babel}

\begin{document}

\title{\textbf{A review of the stochastic background of gravitational waves
in f(R) gravity with WMAP constrains}}

\author{\textbf{Christian Corda}}

\maketitle

\lyxaddress{\begin{center}
Associazione Scientifica Galileo Galilei, Via Pier Cironi 16, I-59100
Prato Italy and 0574news.it - Sezione Scientifico-Tecnologica, via
Sante Pisani 46 - 59100 Prato, Italy
\par\end{center}}

\lyxaddress{\begin{center}
\textit{E-mail address:} \textcolor{blue}{christian.corda@ego-gw.it}
\par\end{center}}
\begin{abstract}
This paper is a review of previous works on the stochastic background
of gravitational waves (SBGWs) which has been discussed in various
peer-reviewed journals and international conferences. The SBGWs is
analyzed with the aid of the Wilkinson Microwave Anisotropy Probe
(WMAP) data. We emphasize that, in general, in previous works in the
literature about the SBGWs, old Cosmic Background Explorer (COBE)
data were used. After this, we want to face the problem of how the
SBGWs and f(R) gravity (where f(R) is a function of the Ricci scalar
R) can be related, showing, vice versa, that a revealed SBGWs could
be a powerful probe for a given theory of gravity. In this way, it
will also be shown that the conform treatment of SBGWs can be used
to parametrize in a natural way f(R) theories. Some interesting examples
which have been recently discussed in the literature will be also
analysed.

The presence and the potential detection of the SBGWs is quite important
in the framework of the debate on high-frequency gravitational waves
(HFGWs) too. Recently, the importance of HFGWs has been emphasized
in some papers in the literature.
\end{abstract}


\section{Introduction}

The accelerated expansion of the Universe, which is today observed,
shows that cosmological dynamic is dominated by the so called Dark
Energy which gives a large negative pressure. This is the standard
picture, in which such new ingredient is considered as a source of
the \textit{right hand side} of the well known Friedman-Robertson-Walker
field equations \cite{key-1}. It should be some form of non-clustered
and non-zero vacuum energy which, together with the clustered Dark
Matter, drives the global dynamics. This is the so called {}``concordance
model'' (Lambda Cold Dark Matter, $\Lambda$CDM) which gives, in
agreement with the Cosmic Microwave Background Radiation (CMBR), Large
Structure Scale (LSS) and Supernovae Ia (SNeIa) data, a good framework
of the today observed Universe, but presents several shortcomings
as the well known {}``coincidence'' and {}``cosmological constant''
problems \cite{key-1}. 

An alternative approach is changing the \textit{left hand side} of
the field equations, seeing if observed cosmic dynamics can be achieved
extending general relativity \cite{key-2,key-3,key-4,key-5}. In this
different context, it is not required to find out candidates for Dark
Energy and Dark Matter, that, till now, have not been found, but only
the {}``observed'' ingredients, which are curvature and baryon matter,
have to be taken into account. Considering this point of view, one
can think that gravity is not scale-invariant and a room for alternative
theories is present \cite{key-6,key-7,key-8}. 

In this picture, even the sensitive detectors for gravitational waves,
like bars and interferometers (i.e. those which are currently in operation
and the ones which are in a phase of planning and proposal stages)
\cite{key-9,key-10,key-11,key-12,key-13,key-14,key-15,key-16}, could,
in principle, be important to confirm or to rule out the physical
consistency of general relativity or of any other theory of gravitation.
This is because, in the context of Extended Theories of Gravity, some
differences between General Relativity and the others theories can
be pointed out starting from the linearized theory of gravity \cite{key-6,key-17,key-18,key-19,key-52}. 

In principle, the most popular Dark Energy and Dark Matter models
can be achieved considering $f(R)$ theories of gravity \cite{key-2,key-3,key-4},
where $R$ is the Ricci curvature scalar. In the tapestry of Extended
Theories of gravity, it is quite important to understand why we are
going to select $f(R)$ theories of gravity among several existing
modified theories of gravity \cite{key-53}. The motivation on such
a choice have been recently well emphasized in the interesting analysis
in \cite{key-54}. Thus, following the review \cite{key-54}, we recall
that some $f(R)$ theories of gravity are not excluded by requirements
of Cosmology and Solar System tests, i.e are \textit{viable}. On the
same time, they have quite rich cosmological structure: they may naturally
describe the effective (cosmological constant, quintessence or phantom)
late-time era with a possible transition from deceleration to acceleration
thanks to gravitational terms which increase with scalar curvature
decrease. In other words, $f(R)$ theories are well tuned with an
accelerating standard FRW universe. These theories may be also rewritten
as scalar-tensor gravity (Einstein frame) with the aid of conform
transformations. Starting by these considerations, in \cite{key-54}
two specific viable models have been considered: the one with $1/R$
and $R^{m}$ terms, and another with $\ln R$ terms. It has been shown
that both of the models may lead to the (cosmological constant or
quintessence) acceleration of the universe as well as an early time
inflation. Moreover, the first model seems to pass the Solar System
tests, i.e. it has the acceptable newtonian limit, no instabilities
and no Brans-Dicke problem (decoupling of scalar) in scalar-tensor
version \cite{key-54}. These versions of $f(R)$ theories may also
describe cosmic speed-up at late universe. 

$f(R)$ theories can be taken into account also in the analysis of
the SBGWs which, together with the CMBR, would carry, if detected,
a huge amount of information on the early stages of the Universe evolution
\cite{key-20,key-21,key-52}). A key role for the production and the
detection of this graviton background is played by the adopted theory
of gravity \cite{key-21,key-52} in this case too. 

In the second section of this review, the SBGWs is analysed with the
aid of the WMAP data \cite{key-20,key-22,key-23}. We emphasize that,
in general, in previous works in the literature about the SBGWs, old
COBE data were used (see \cite{key-24,key-25,key-26,key-27,key-28,key-29}
for example).

In the third section we want to face the problem of how the SBGWs
and $f(R)$ gravity can be related, showing, vice versa, that a revealed
SBGWs could be a powerful probe for a given theory of gravity. In
this way, it will also be shown that the conform treatment of SBGWs
can be used to parametrize in a natural way f(R) theories. In this
framework, we will discuss like particular case \cite{key-53} some
interesting viable $f(R)$ theories that have been analysed in refs.
\cite{key-5,key-55,key-56}. In this way, a discussion which started
in \cite{key-57} will be integrated.

The presence and the potential detection of the SBGWs is quite important
in the framework of the debate on high-frequency gravitational waves
(HFGWs) too. Recently, the importance of HFGWs has been emphasized
in some papers in the literature \cite{key-48,key-49,key-50,key-51}. 

We emphasize that this is a review of previous works on the SBGWs
which have been discussed in various peer-reviewed journals and international
conferences.

\section{Constrains on the stochastic background of gravitational waves from
the WMAP data}

From our analysis, it will result that the WMAP bounds on the energy
spectrum and on the characteristic amplitude of the SBGWs are greater
than the COBE ones, but they are also far below frequencies of the
earth-based antennas band. At the end of this section a lower bound
for the integration time of a potential detection with advanced LIGO
is released and compared with the previous one arising from the old
COBE data. Even if the new lower bound is minor than the previous
one, it results very long, thus for a possible detection we hope in
the Laser Interferometric Space Antenna (LISA) and in a further growth
in the sensitivity of advanced projects.

The strongest constraint on the spectrum of the relic SBGWs in the
frequency range of ground based antennas like bars and interferometers,
which is the range $10Hz\leq f\leq10^{4}Hz$, comes from the high
isotropy observed in the CMBR.

The fluctuation $\Delta T$ of the temperature of CMBR from its mean
value $T_{0}=2.728$ K varies from point to point in the sky \cite{key-20,key-22,key-23},
and, since the sky can be considered the surface of a sphere, the
fitting of $\Delta T$ is performed in terms of a Laplace series of
spherical harmonics

\begin{equation}
\frac{\Delta T}{T_{0}}(\hat{\Omega})=\sum_{l=1}^{\infty}\sum_{m=-l}^{l}a_{lm}Y_{lm}(\hat{\Omega})\label{eq: fluttuazioni CBR}\end{equation}

and the fluctuations are assumed statistically independent ($<a_{lm}>=0$,
$<a_{lm}a_{l'm'}^{*}>=C_{l}\delta_{ll'}\delta_{mm'}$). $\hat{\Omega}$
denotes a point on the 2-sphere in Eq. (\ref{eq: fluttuazioni CBR}),
while the $a_{lm}$ are the multipole moments of the CMBR. For details
about the definition of statistically independent fluctuations in
the context of the temperature fluctuations of CMBR see \cite{key-24,key-25}.

The WMAP data \cite{key-22,key-23} permit a more precise determination
of the rms quadrupole moment of the fluctuations than the COBE data

\begin{equation}
Q_{rms}\equiv T(\sum_{m=-2}^{2}\frac{|a_{2m}|^{2}}{4\pi})^{\frac{1}{2}}=8\pm2\mu K,\label{eq: Q rms}\end{equation}

while in the COBE data we had \cite{key-25,key-26,key-27}

\begin{equation}
Q_{rms}=14.3_{-3.3}^{+5.2}\mu K.\label{eq: Q rms COBE}\end{equation}
 A connection between the fluctuation of the temperature of the CMBR
and the SBGWs derives from the \textit{Sachs-Wolfe effect} \cite{key-20,key-30}.
Sachs and Wolfe showed that variations in the density of cosmological
fluid and GWs perturbations result in the fluctuation of the temperature
of the CMBR, even if the surface of last scattering had a perfectly
uniform temperature \cite{key-30}. In particular, the fluctuation
of the temperature (at the lowest order) in a particular point of
the space is

\begin{equation}
\frac{\Delta T}{T_{0}}(\hat{\Omega})=\frac{1}{2}\int_{nullgeodesic}d\lambda\frac{\partial}{\partial\eta}h_{rr}.\label{eq: null geodesic}\end{equation}

The integral in Eq. (\ref{eq: null geodesic}) is taken over a path
of null geodesic which leaves the current spacetime point heading
off in the direction defined by the unit vector $\hat{\Omega}$ and
going back to the surface of last scattering of the CMBR.

Here $\lambda$ is a particular choice of the affine parameter along
the null geodesic. By using conform coordinates, we have for the metric
perturbation

\begin{equation}
\delta g_{ab}=R^{2}(\eta)h_{ab}\label{eq: metric perturbation}\end{equation}

and $r$ in Eq. (\ref{eq: null geodesic}) is a radial spatial coordinate
which goes outwards from the current spacetime point. The effect of
a long wavelength GW is to shift the temperature distribution of CMBR
from perfect isotropy. As the fluctuations are very small ($<\Delta T/T_{0}>\leq5\times10^{-5}$
\cite{key-22,key-23}), the perturbations caused by the relic SBGWs
cannot be too large. 

The WMAP results give rather tight constraints on the spectrum of
the SBGWs at very long wavelengths. In \cite{key-24,key-25} we find
a constraint on $\Omega_{gw}(f)$ which derives from the COBE observational
limits, given by

\begin{equation}
\Omega_{gw}(f)h_{100}^{2}<7\times10^{-11}(\frac{H_{0}}{f})^{2}\textrm{ for }H_{0}<f<30H_{0}.\label{eq: limite COBE}\end{equation}

We included a dimensionless factor $h_{100}$ which now is just a
convenience (in the past it came from an uncertainty in the value
of $H_{0}$). From the WMAP data we have $h_{100}=0.72\pm0.05$ \cite{key-22,key-23}.

Now, the same constraint will be obtained from the WMAP data \cite{key-20}.
Because of its specific polarization properties, relic SBGWs should
generate particular polarization patterns of CMBR anisotropies, but
the detection of CMBR polarizations is not fulfilled today \cite{key-31}.
Thus, an indirect method will be used. We know that relic GWs have
very long wavelengths of Hubble radius size, then the CMBR power spectrum
from them should manifest fast decrease at smaller scales (high multipole
moments). But we also know that scalar modes produce a rich CMBR power
spectrum at large multipole moments (series of acoustic peaks, \cite{key-22,key-23}).
Then, the properties of tensor modes in the cosmological perturbations
of spacetime metric can be extract from observational data using angular
CMBR power spectrum combined with large scale structure of the Universe.
One can see (Fig. 1 ) that in the range $2\leq l\leq30$ (the same
used in \cite{key-25}, but with the old COBE data \cite{key-32})
scalar and tensor contributions are comparable. 

\begin{figure}
\includegraphics[clip,scale=0.9]{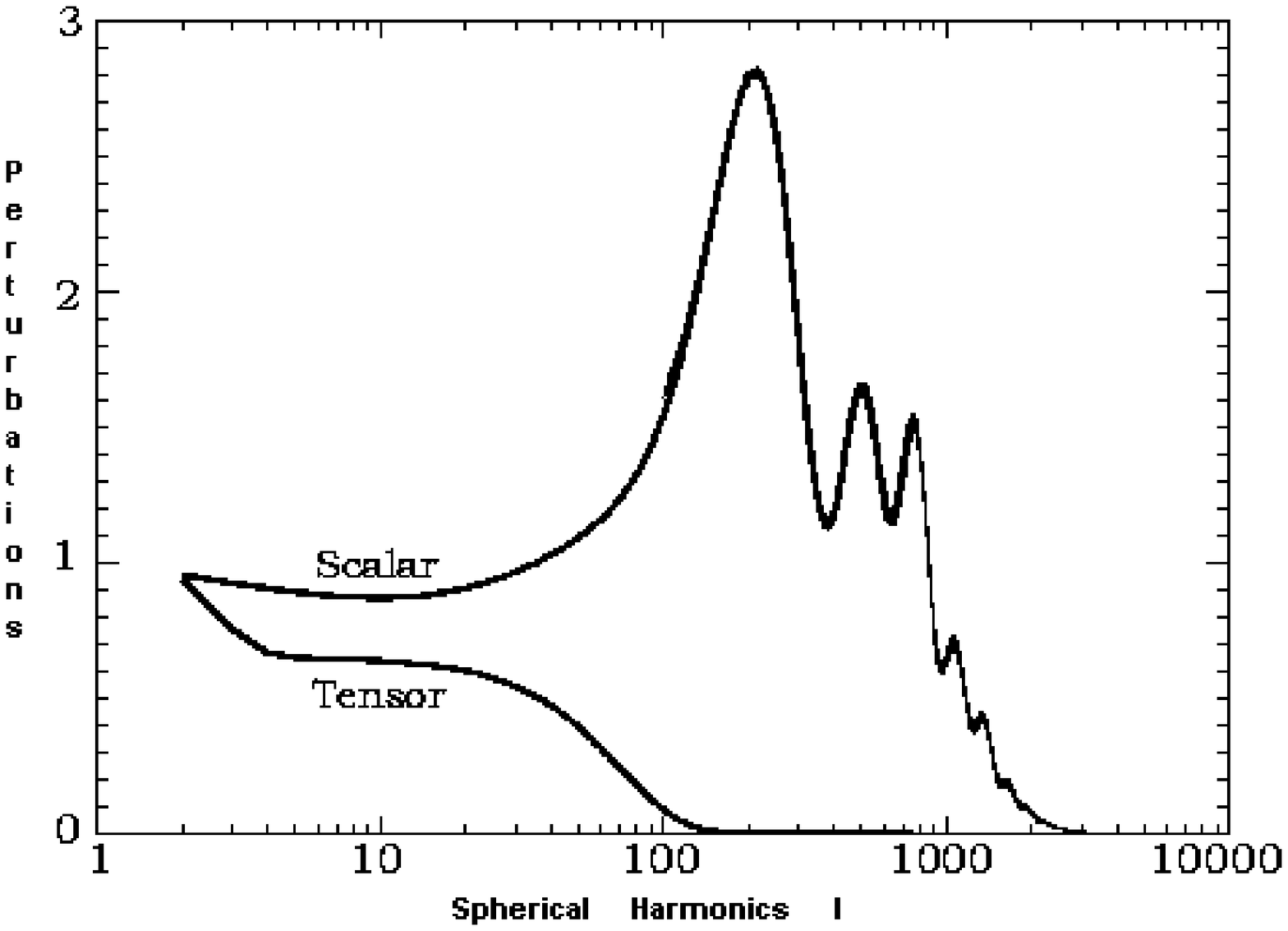}

\textbf{Figure 1: the tensor to scalar ratio, adapted from C. Corda}
\textbf{Mod. Phys. Lett. A 22, 16, 1167-1173 (2007)}
\end{figure}

From \cite{key-22,key-23}, the WMAP data give for the tensor/scalar
ratio $r$ the constraint $r<0.9$. \foreignlanguage{italian}{(}$r<0.5$
\foreignlanguage{italian}{in the COBE data} \cite{key-32}\foreignlanguage{italian}{);
Novosyadly and Apunevych} obtained $\Omega_{scalar}(H_{0})<2.7\times10^{-9}$
\cite{key-31}. Thus, if one remembers that, at order of Hubble radius,
the tensor spectral index is $-4\leq n_{t}\leq-2$ \cite{key-20},
it results

\begin{equation}
\Omega_{gw}(f)h_{100}^{2}<1.6\times10^{-9}(\frac{H_{0}}{f})^{2}\textrm{ for }H_{0}<f<30H_{0},\label{eq: limite WMAP}\end{equation}

which is greater than the COBE data result of eq. (\ref{eq: limite COBE}). 

We emphasize that the limit of Eq. (\ref{eq: limite WMAP}) is not
a limit for any GWs, but only for relic ones of cosmological origin,
which were present at the era of the CMBR decoupling. Also, the same
limit only applies over very long wavelengths (i.e. very low frequencies)
and it is far below frequencies of the Virgo - LIGO (Laser Interferometric
Gravitational Observatory) band.

The primordial production of the relic SBGWs has been analysed in
\cite{key-25,key-26,key-27,key-33} (note: a generalization for f(R)
theories of gravity will be given in the next section following \cite{key-21}),
where it has been shown that in the range $10^{-15}Hz\leq f\leq10^{10}Hz$
the spectrum is flat and proportional to the ratio

\begin{equation}
\frac{\rho_{ds}}{\rho_{Planck}}\approx10^{-12}.\label{eq: rapporto densita' primordiali}\end{equation}

As the spectrum remains flat at high frequency, the relic SBGWs are
quite important also in the framework of HFGWs, which have been emphasized
in some works in the recent literature \cite{key-48,key-49,key-50,key-51}.

WMAP observations put strongly severe restrictions on the spectrum,
as we discussed above. In Fig. 2 the spectrum $\Omega_{gw}$ is mapped:
the amplitude (determined by the ratio $\frac{\rho_{ds}}{\rho_{Planck}}$)
has been chosen to be \textit{as large as possible, consistent with
the WMAP constraint} (\ref{eq: limite WMAP}). 

\begin{figure}
\includegraphics{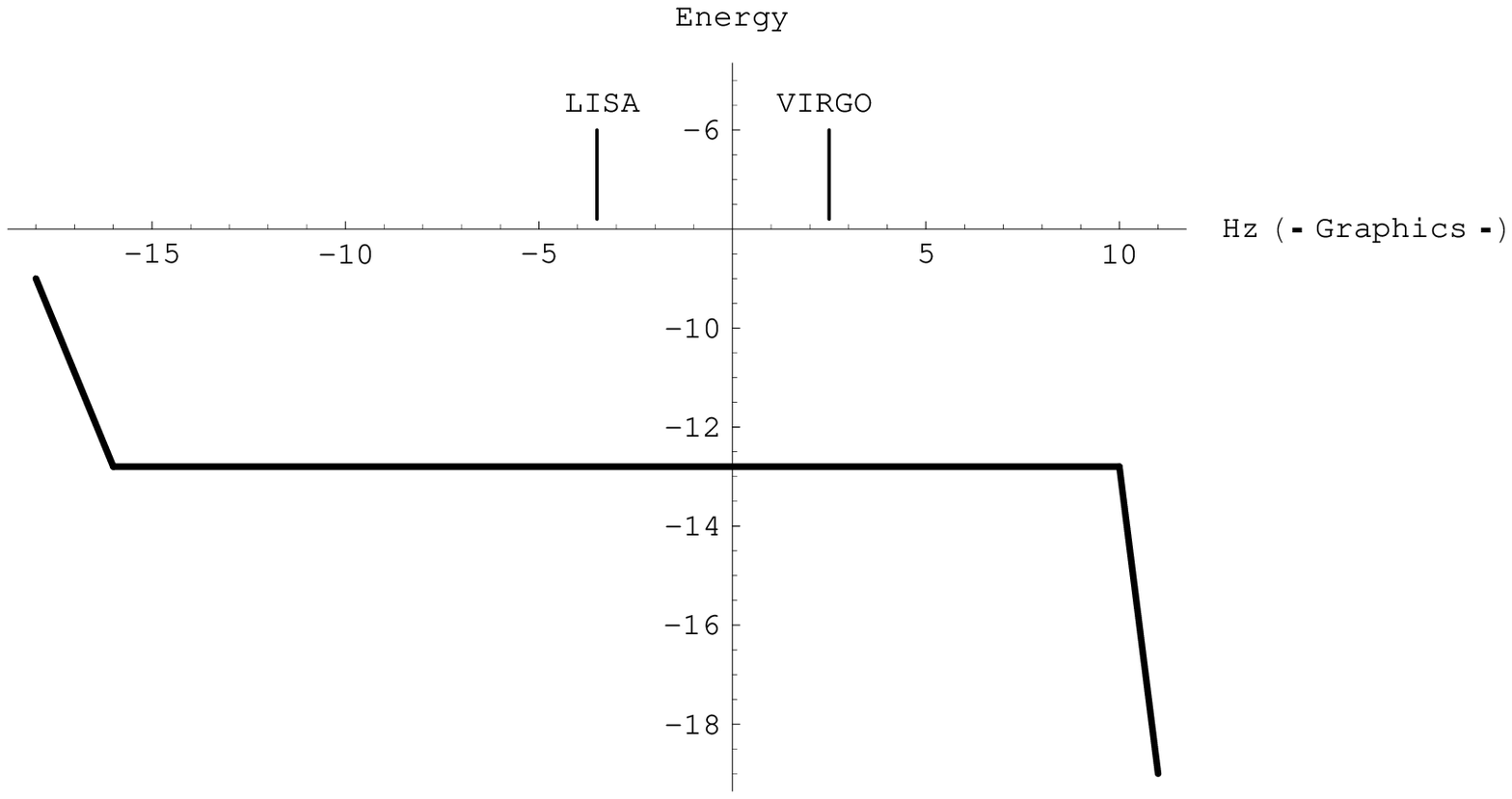}

\textbf{Figure 2: the WMAP bound, adapted from C. Corda - Astropart.
Phys., 30, 209-215 (2008)}

\end{figure}

Nevertheless, because the spectrum falls off $\propto f^{-2}$ at
low frequencies \cite{key-25,key-26,key-27,key-33}, this means that
today, at Virgo and LISA frequencies, indicated in Fig. 2,

\begin{center}
\begin{equation}
\Omega_{gw}(f)h_{100}^{2}<9\times10^{-13},\label{eq: limite spettro WMAP}\end{equation}

\par\end{center}

while using the COBE data it was

\begin{center}
$\Omega_{gw}(f)h_{100}^{2}<8\times10^{-14}$(refs. \cite{key-25,key-32}).
\par\end{center}

It is interesting to calculate the correspondent strain at $\approx100Hz$,
where interferometers like Virgo and LIGO have a maximum in sensitivity.
The well known equation for the characteristic amplitude \cite{key-20,key-24,key-25,key-29}
can be used:

\begin{equation}
h_{c}(f)\simeq1.26\times10^{-18}(\frac{1Hz}{f})\sqrt{h_{100}^{2}\Omega_{gw}(f)},\label{eq: legame ampiezza-spettro}\end{equation}
obtaining

\begin{equation}
h_{c}(100Hz)<1.7\times10^{-26}.\label{eq: limite per lo strain}\end{equation}

Then, as for ground-based interferometers a sensitivity of the order
of $10^{-22}$ is expected at $\approx100Hz$, four order of magnitude
have to be gained in the signal to noise ratio \cite{key-9,key-10,key-11,key-12,key-13,key-14,key-15,key-16}.
Let us analyze smaller frequencies too. The sensitivity of the Virgo
interferometer is of the order of $10^{-21}$ at $\approx10Hz$ \cite{key-9,key-10}
and in that case it is 

\begin{equation}
h_{c}(10Hz)<1.7\times10^{-25}.\label{eq: limite per lo strain2}\end{equation}

For a better understanding of the difficulties on the detection of
the SBGWs, a lower bound for the integration time of a potential detection
with advanced LIGO is released. For a cross-correlation between two
interferometers the signal to noise ratio (SNR) increases as 

\begin{equation}
SNR=\sqrt{2T}\frac{H_{0}^{2}}{5\pi^{2}}\sqrt{\int_{0}^{\infty}df\frac{\Omega_{gw}^{2}(f)\gamma^{2}(f)}{f^{6}P_{1}(|f|)P_{2}(|f|)}},\label{eq: SNR2}\end{equation}

where $P_{i}(|f|)$ is the one-sided power spectral density of the
$i$ detector \cite{key-34} and $\gamma(f)$ the well known overlap-reduction
function \cite{key-34,key-35}. Assuming two coincident co-aligned
detectors $(\gamma(f)=1)$ with a noise of order $10^{-48}/Hz$ (i.e.
a typical value for the advanced LIGO sensitivity \cite{key-36}),
one gets $(SNR)\sim1$ for $\sim3*\times10^{5}years$ using our result
$\Omega_{gw}(f)h_{100}^{2}\sim9\times10^{-13}$ while it is $(SNR)\sim1$
for $\sim3\times10^{7}years$ using previous COBE result $\Omega_{gw}(f)h_{100}^{2}\sim8\times10^{-14}$.
As the overlap reduction function degrades the SNR, these results
can be considered a solid upper limit for the advanced LIGO configuration
for the two different values of the spectrum. 

The sensitivity of the LISA interferometer will be of the order of
$10^{-22}$ at $10^{-3}Hz$ \cite{key-37} and in that case it is 

\begin{equation}
h_{c}(10^{-3}Hz)<1.7\times10^{-21}.\label{eq: limite per lo strain3}\end{equation}

Then, a stochastic background of relic gravitational waves could be
in principle detected by the LISA interferometer. We also hope in
a further growth in the sensitivity of advanced projects.

We emphasize that the assumption that all the tensor perturbation
in the Universe are due to a SBGWs is quite strong, but our results
(\ref{eq: limite spettro WMAP}), (\ref{eq: limite per lo strain}),
(\ref{eq: limite per lo strain2}) and (\ref{eq: limite per lo strain3})
can be considered like upper bounds. 

Resuming, in this section the SBGWs has been analysed with the aid
of the WMAP data, while previous works in the literature used the
old COBE data, seeing that the predicted signal for these relic GWs
is very weak. From our analysis it resulted that the WMAP bound on
the energy spectrum and on the characteristic amplitude of the SBGWs
are greater than the COBE ones, but they are also far below frequencies
of the earth-based antennas band. In fact, the integration time of
a potential detection with advanced interferometers is very long,
thus, for a possible detection, we have to hope in a further growth
in the sensitivity of advanced ground based projects and in the LISA
interferometer.

\section{Stochastic background and f(R) theories of gravity}

GWs are the perturbations $h_{\alpha\beta}$ of the metric $g_{\alpha\beta}$
which transform as three-tensors. Following \cite{key-21,key-38},
the GW-equations in the TT gauge are 

\begin{equation}
\square h_{i}^{j}=0,\label{eq: dalembert}\end{equation}
where $\square\equiv(-g)^{-1/2}\partial_{\alpha}(-g)^{1/2}g^{\alpha\beta}\partial_{\beta}$
is the usual d'Alembert operator and these equations are derived from
the Einstein field equations deduced from the Einstein-Hilbert Lagrangian
density $L=R$ \cite{key-6,key-21}. Clearly, matter perturbations
do not appear in (\ref{eq: dalembert}) since scalar and tensor perturbations
do not couple with tensor perturbations in Einstein equations. The
Latin indices run from 1 to 3, the Greek ones from 0 to 3. Our task
is now to derive the analogous of Eqs. (\ref{eq: dalembert}) assuming
a generic theory of gravity given by the action \begin{equation}
A=\frac{1}{2k}\int d^{4}x\sqrt{-g}f(R),\label{eq: high order}\end{equation}

where, for a sake of simplicity, we have discarded matter contributions.
A conform analysis will help in this goal. In fact, assuming the conform
transformation \begin{equation}
\tilde{g}_{\alpha\beta}=e^{2\Phi}g_{\alpha\beta}\label{eq: conforme}\end{equation}

where the conform rescaling \begin{equation}
e^{2\Phi}=|f'(R)|\label{eq: rescaling}\end{equation}

has been chosen being the prime the derivative in respect to the Ricci
curvature scalar and $\Phi$ the {}``conform scalar field'', we
obtain the conform equivalent Hilbert-Einstein action \begin{equation}
A=\frac{1}{2k}\int d^{4}x\sqrt{-\widetilde{g}}[\widetilde{R}+L(\Phi,\Phi_{;\alpha})],\label{eq: conform}\end{equation}

where $L(\Phi,\Phi_{;\alpha})$ is the conform scalar field contribution
derived from

\begin{equation}
\tilde{R}_{\alpha\beta}=R_{\alpha\beta}+2(\Phi_{;\alpha}\Phi_{;\beta}-g_{\alpha\beta}\Phi_{;\delta}\Phi^{;\delta}-\frac{1}{2}g_{\alpha\beta}\Phi^{;\delta}{}_{;\delta})\label{eq: conformRicci}\end{equation}

and \begin{equation}
\tilde{R}=e^{-2\Phi}+(R-6\square\Phi-6\Phi_{;\delta}\Phi^{;\delta}).\label{eq: conformRicciScalar}\end{equation}

In any case, as we will see, the $L(\Phi,\Phi_{;\alpha})$-term does
not affect the GWs-tensor equations, thus it will not be considered
any longer (note: a scalar component in GWs is often considered \cite{key-17,key-39,key-40,key-41,key-52},
but here we are taking into account only the genuine tensor part of
stochastic background).

Starting from the action (\ref{eq: conform}) and deriving the Einstein-like
conform equations, the GWs equations are \begin{equation}
\widetilde{\square}\widetilde{h}_{i}^{j}=0,\label{eq: dalembert conf}\end{equation}

expressed in the conform metric $\tilde{g}_{\alpha\beta}.$ As no
scalar perturbation couples to the tensor part of gravitational waves,
it is 

\begin{equation}
\widetilde{h}_{i}^{j}=\widetilde{g}^{lj}\delta\widetilde{g}_{il}=e^{-2\Phi}g^{lj}e^{2\Phi}\delta g_{il}=h_{i}^{j},\label{eq: confinvariant}\end{equation}

which means that $h_{i}^{j}$ is a conform invariant.

As a consequence, the plane wave amplitude $h_{i}^{j}=h(t)e_{i}^{j}\exp(ik_{i}x^{j}),$
where $e_{i}^{j}$ is the polarization tensor, are the same in both
metrics. In any case the d'Alembert operator transforms as 

\begin{equation}
\widetilde{\square}=e^{-2\Phi}(\square+2\Phi^{;\alpha}\partial_{;\alpha})\label{eq: quadratello}\end{equation}

and this means that the background is changing while the tensor wave
amplitude is fixed. 

In order to study the cosmological stochastic background, the operator
(\ref{eq: quadratello}) can be specified for a Friedman-Robertson-Walker
metric \cite{key-26,key-27}, and Eq. (\ref{eq: dalembert conf})
becomes

\begin{equation}
\ddot{h}+(3H+2\dot{\Phi})\dot{h}+k^{2}a^{-2}h=0,\label{eq: evoluzione h}\end{equation}

being $\square=\frac{\partial}{\partial t^{2}}+3H\frac{\partial}{\partial t}$,
$a(t)$ the scale factor and $k$ the wave number. It has to be emphasized
that Eq. (\ref{eq: evoluzione h}) applies to any f(R) theory whose
conform transformation can be defined as $e^{2\Phi}=f'(R).$ The solution,
i.e. the GW amplitude, depends on the specific cosmological background
(i.e. $a(t)$) and the specific theory of gravity. For example, assuming
power law behaviors for $a(t)$ and $\Phi(t)=\frac{1}{2}\ln f'(R(t)),$
that is

\begin{equation}
\Phi(t)=f'(R)=f_{0}'(\frac{t}{t_{0}})^{m},\textrm{ $ $$ $$ $$ $$ $$ $$ $$ $ }a(t)=a_{0}(\frac{t}{t_{0}})^{n},\label{eq: phi a}\end{equation}

it is easy to show that general relativity is recovered for $m=0$
while 

\begin{equation}
n=\frac{m^{2}+m-2}{m+2}\label{eq: m n}\end{equation}

is the relation between the parameters for a generic $f(R)=f_{0}R^{s}$
where $s=1-\frac{m}{2}$ with $s\neq1$ \cite{key-42}. Equation (\ref{eq: evoluzione h})
can be recast in the form \begin{equation}
\ddot{h}+(3n+m)t^{-1}\dot{h}+k^{2}a_{0}(\frac{t_{0}}{t})^{2n}h=0,\label{eq: evoluzione h 2}\end{equation}

whose general solution is \begin{equation}
h(t)=(\frac{t_{0}}{t})^{-\beta}[C_{1}J_{\alpha}(x)+C_{2}J_{-\alpha}(x)].\label{eq: sol ev h}\end{equation}

$J_{\alpha}$'s are Bessel functions and

\begin{equation}
\alpha=\frac{1-3n-m}{2(n-1)},\textrm{ $ $$ $$ $}\beta=\frac{1-3n-m}{2},\textrm{ $ $$ $$ $}x=\frac{kt^{1-n}}{1-n},\label{eq: alfabetx}\end{equation}

while $t_{0},$ $C_{1},$ and $C_{2}$ are constants related to the
specific values of $n$ and $m.$

The time units are in terms of the Hubble radius $H^{-1};$ $n=1/2$
is a radiation-like evolution; $n=2/3$ is a dust-like evolution,
$n=2$ labels power-law inflationary phases and $n=-5$ is a pole-like
inflation. From Eq. (\ref{eq: m n}), a singular case is for $m=-2$
and $s=2.$ It is clear that the conform invariant plane-wave amplitude
evolution of the tensor GW strictly depends on the background.

Now, let us take into account the issue of the production of relic
GWs contributing to the stochastic background. Several mechanisms
can be considered as cosmological populations of astrophysical sources
\cite{key-43}, vacuum fluctuations, phase transitions \cite{key-24}
and so on. In principle, we could seek for contributions due to every
high-energy physical process in the early phases of the Universe evolution.

It is important to distinguish processes coming from transitions like
inflation, where the Hubble flow emerges in the radiation dominated
phase and processes, like the early star formation rates, where the
production takes place during the dust dominated era. In the first
case, stochastic GWs background is strictly related to the cosmological
model. This is the case we are considering here which is, furthermore,
also connected to the specific theory of gravity. In particular, one
can assume that the main contribution to the stochastic background
comes from the amplification of vacuum fluctuations at the transition
between an inflationary phase and the radiation-dominated era. However,
in any inflationary model, we can assume that the relic GWs generated
as zero-point fluctuations during the inflation undergoes adiabatically
damped oscillations ($\sim1/a$) until they reach the Hubble radius
$H^{-1}.$ This is the particle horizon for the growth of perturbations.
On the other hand, any other previous fluctuation is smoothed away
by the inflationary expansion. The GWs freeze out for $a/k\gg H^{-1}$
and re-enter the $H^{-1}$ radius after the reheating in the Friedman
era \cite{key-21,key-25,key-26,key-27,key-33}. Re-entering in the
radiation dominated or in the dust-dominated era depends on the scale
of the GW. After the re-entering, GWs can be detected by their Sachs-Wolfe
effect on the temperature anisotropy $\frac{\bigtriangleup T}{T}$
at the decoupling \cite{key-30}. When $\Phi$ acts as the inflaton
\cite{key-21,key-44} we have $\dot{\Phi}\ll H$ during the inflation.
Considering the conformal time $d\eta=dt/a$, Eq. (\ref{eq: evoluzione h})
reads\begin{equation}
h''+2\frac{\gamma'}{\gamma}h'+k^{2}h=0,\label{eq: evoluzione h 3}\end{equation}

where $\gamma=ae^{\Phi}$ and derivation is with respect to $\eta.$
Inflation means that $a(t)=a_{0}\exp(Ht)$ and then $\eta=\int dt/a=1/(aH)$
and $\frac{\gamma'}{\gamma}=-\frac{1}{\eta}.$ The exact solution
of (\ref{eq: evoluzione h 3}) is \begin{equation}
h(\eta)=k{}^{-3/2}\sqrt{2/k}[C_{1}(\sin k\eta-\cos k\eta)+C_{2}(\sin k\eta+\cos k\eta)].\label{eq: sol ev h2}\end{equation}

Inside the $1/H$ radius it is $k\eta\gg1.$ Furthermore, considering
the absence of gravitons in the initial vacuum state, we have only
negative-frequency modes and then the adiabatic behavior is \begin{equation}
h=k{}^{1/2}\sqrt{2/\pi}\frac{1}{aH}C\exp(-ik\eta).\label{eq: sol ev h3}\end{equation}

At the first horizon crossing ($aH=k$), the averaged amplitude $A_{h}=(k/2\pi)^{3/2}|h|$
of the perturbations is \begin{equation}
A_{h}=\frac{1}{2\pi{}^{2}}C.\label{eq: Ah}\end{equation}

When the scale $a/k$ grows larger than the Hubble radius $1/H,$
the growing mode of evolution is constant, that it is frozen. This
situation corresponds to the limit $-k\eta\ll1$ in Eq. (\ref{eq: sol ev h2}).
Since $\Phi$ acts as the inflaton field, it is $\Phi\sim0$ at re-entering
(after the end of inflation). Then, the amplitude $A_{h}$ of the
wave is preserved until the second horizon crossing after which it
can be observed, in principle, as an anisotropy perturbation of the
CBR. It can be shown that $\frac{\bigtriangleup T}{T}\lesssim A_{h}$
is an upper limit to $A_{h}$ since other effects can contribute to
the background anisotropy \cite{key-45}. From this consideration,
it is clear that the only relevant quantity is the initial amplitude
$C$ in equation (\ref{eq: sol ev h3}) which is conserved until the
re-entering. Such an amplitude directly depends on the fundamental
mechanism generating perturbations. Inflation gives rise to processes
capable of producing perturbations as zero-point energy fluctuations.
Such a mechanism depends on the adopted theory of gravitation and
then $\frac{\bigtriangleup T}{T}$ could constitute a further constraint
to select a suitable f(R)-theory. Considering a single graviton in
the form of a monochromatic wave, its zero-point amplitude is derived
through the commutation relations \begin{equation}
[h(t,x),\pi_{h}(t,y)]=i\delta^{3}(x-y),\label{eq: commutare}\end{equation}

calculated at a fixed time $t,$ where the amplitude $h$ is the field
and $\pi_{h}$ is the conjugate momentum operator. Writing the Lagrangian
for $h$ 

\begin{equation}
\widetilde{L}=\frac{1}{2}\sqrt{\widetilde{g}}\widetilde{g}^{\alpha\beta}h_{;\alpha}h_{;\beta}\label{eq: lagrange}\end{equation}

in the conformal FRW metric $\widetilde{g}_{\alpha\beta}$ ($h$ is
a conform invariant), we obtain

\begin{equation}
\pi_{h}=\frac{\partial\widetilde{L}}{\partial\dot{h}}=e^{2\Phi}a^{3}\dot{h}.\label{eq: pi h}\end{equation}

Equation (\ref{eq: commutare}) becomes\begin{equation}
[h(t,x),\dot{h}(y,y)]=i\frac{\delta^{3}(x-y)}{e^{2\Phi}a^{3}}\label{eq: commutare 2}\end{equation}

and the fields $h$ and $\dot{h}$ can be expanded in terms of creation
and annihilation operators

\begin{equation}
h(t,x)=\frac{1}{(2\pi)^{3/2}}\int d^{3}k[h(t)e^{-ikx}+h^{*}(t)e^{ikx}],\label{eq: crea}\end{equation}
\begin{equation}
\dot{h}(t,x)=\frac{1}{(2\pi)^{3/2}}\int d^{3}k[\dot{h}(t)e^{-ikx}+\dot{h}^{*}(t)e^{ikx}].\label{eq: distruggi}\end{equation}

The commutation relations in conform time are then \begin{equation}
[h(h'^{*}-h^{*}h']=i\frac{(2\pi)^{3}}{e^{2\Phi}a^{3}}.\label{eq: commutare 3}\end{equation}

Putting (\ref{eq: sol ev h3}) and (\ref{eq: Ah}), we obtain $C=\sqrt{2}\pi{}^{2}He^{-\Phi}$
where $H$ and $\Phi$ are calculated at the first horizon crossing
and then \begin{equation}
A_{h}=\frac{\sqrt{2}}{2}{}He^{-\Phi},\label{eq: Ah2}\end{equation}

which means that the amplitude of GWs produced during inflation directly
depends on the given f(R) theory being $\Phi=\frac{1}{2}\ln|f'(R)|.$
Explicitly, it is \begin{equation}
A_{h}=\frac{H}{\sqrt{2|f'(R)|}}.\label{eq: Ah3}\end{equation}
 This result deserves some discussion and can be read in two ways.
From one side the amplitude of relic GWs produced during inflation
depends on the given theory of gravity that, if different from general
relativity, gives extra degrees of freedom which assume the role of
inflaton field in the cosmological dynamics \cite{key-44}. On the
other hand, the Sachs-Wolfe effect related to the CMBR temperature
anisotropy could constitute a powerful tool to test the true theory
of gravity at early epochs, i.e. at very high redshift. This probe,
related with data a medium \cite{key-46} and low redshift \cite{key-47},
could strongly contribute to reconstruct cosmological dynamics at
every scale, to further test general relativity or to rule out it
against alternative theories and to give constrains on the SBGWs,
if f(R) theories ares independently probed at other scales. 

For a better understanding of the issue, it is important to analyse
in spirit of our discussion some particular viable model \cite{key-53},
for example the interesting viable models in \cite{key-5,key-55,key-56}.
In this way we will also integrate an analysis which started in \cite{key-57}. 

Let us start by the model proposed in \cite{key-56,key-57}. In this
case it is

\begin{equation}
f(R)=\frac{\alpha R^{2n}-\beta R^{n}}{1+\gamma R^{n}},\label{eq: f1}\end{equation}

which, in our case, gives the conform scalar field 

\begin{equation}
\Phi=\frac{1}{2}\ln|\frac{nR^{n-1}(\alpha\gamma R^{2n}-2\alpha R^{n}-\beta}{(1+\gamma R^{n})^{2}}|.\label{eq: scal 1}\end{equation}

Here $\alpha,$ $\beta,$ $\gamma$ are positive constants and $n$
is a positive integer \cite{key-56,key-57}. These equations cannot
satisfy the curvature $R_{0}$ in the present Universe for which it
is $f(R_{0})=0$ because this implies a singularity in the amplitude
of the relic GWs (\ref{Ah4}). 

In the model proposed in \cite{key-5} it is \begin{equation}
f(R)=-\alpha[\tanh\frac{b(R-R_{0})}{2}+\tanh\frac{bR_{0}}{2}]\label{eq: f2}\end{equation}

and\begin{equation}
\Phi=\frac{1}{2}\ln|\frac{\alpha b}{2\cosh^{2}\frac{b(R-R_{0})}{2}}|.\label{eq: scal 2}\end{equation}

Here $\alpha,$ and $b$ are positive constants \cite{key-5,key-57}.
Thus, in the case of flat Universe $R=0$, it is $f(0)=0$ and \begin{equation}
A_{h}=H\sqrt{|\frac{\cosh^{2}\frac{b(R-R_{0})}{2}}{\alpha b}|}.\label{Ah4}\end{equation}

On the other hand, in the case in which it is $R\ggg R_{0}$ in the
present Universe it is \cite{key-5,key-57}

\begin{equation}
f(R)\rightarrow2\Lambda_{eff}\equiv-\alpha[1+\tanh\frac{bR_{0}}{2}],\label{eq: f3}\end{equation}

i.e. $\Lambda_{eff}$ plays the role of an effective cosmological
constant. In this case $A_{h}$ increases with increasing curvature. 

$A_{h}$ has a maximum which is\begin{equation}
A_{h}=H\sqrt{|\frac{-2}{\alpha b}|}\label{eq: Max}\end{equation}

in the case in which $R=R_{0}.$  

Two pretty realistic f(R) theories have been proposed in \cite{key-55}.
The first is \begin{equation}
f(R)=\frac{(R-R_{0})^{2n+1}+R_{0}^{2n+1}}{f_{0}+f_{1}\{(R-R_{0})^{2n+1}+R_{0}^{2n+1}\}},\label{:eq: f4}\end{equation}

which gives the correspondent conform scalar field\begin{equation}
\Phi=\frac{1}{2}\ln|\frac{(2n+1)f_{0}(R-R_{0})^{2n}}{(f_{0}+f_{1}\{(R-R_{0})^{2n+1}+R_{0}^{2n+1}\})^{2}}|.\label{eq: scal 3}\end{equation}

Here $n$ is a positive integer $f_{1}=1/\Lambda_{i},$ where $\Lambda_{i}$
represents the effective cosmological constant and $f_{0}\sim\frac{1}{2}R_{0}^{2n}$
\cite{key-55}. In this case the amplitude of the stochastic background
admits a minimum at the point $R=\tilde{R},$ where $\tilde{R}$ satisfies

\begin{equation}
(\tilde{R}-R_{0})^{2n+1}=\frac{f_{0}+f_{1}R_{0}{}^{2n+1}}{(n+1)f_{1}}\label{eq: soddisfa}\end{equation}

and the minimum is given by \begin{equation}
A_{h}\sim H\sqrt{\frac{(f_{1}R_{0})^{\frac{2n}{2n+1}}}{2}}.\label{:eq: Ah Min}\end{equation}

The second proposal in \cite{key-55} is the theory \begin{equation}
f(R)=-f_{0}\intop_{0}^{R}dR\exp(-\alpha\frac{R_{1}{}^{2n}}{(R-R_{1})^{2n}}-\frac{R}{\beta\Lambda_{i}}),\label{eq: f5}\end{equation}

which gives the correspondent conform scalar field \begin{equation}
\Phi=\frac{1}{2}(-\alpha\frac{R_{1}{}^{2n}}{(R-R_{1})^{2n}}-\frac{R}{\beta\Lambda_{i}})+\frac{1}{2}\ln f_{0}.\label{eq: scal 4}\end{equation}

Here $\alpha,$ $\beta,$ $f_{0}$ and $R_{1}$are constants, while
$f(+\infty)\sim f_{0}\beta\Lambda_{i}$ is identified with the effective
cosmological constant at the inflation era if $f_{0}\beta=1.$ The
case of flat Universe is interesting. We get\begin{equation}
A_{h}\sim H\sqrt{|\frac{\exp(\alpha R_{1}^{2n-1})}{2f_{0}}|}.\label{eq: Ah piatto}\end{equation}

Resuming, in this section it has been shown that amplitudes of tensor
GWs are conform invariant and their evolution depends ond the cosmological
SBGWs. Such a background is tuned by a conform scalar field which
is not present in the standard general relativity. Assuming that primordial
vacuum fluctuations produce a SBGWs, beside scalar perturbations,
kinetic distortions and so on, the initial amplitude of these ones
is function of the f(R)-theory of gravity and then the SBGWs can be,
in a certain sense, {}``tuned'' by the theory. Vice versa, data
coming from the Sachs-Wolfe effect could contribute to select a suitable
f(R)-theory which can be consistently matched with other observations. 

We have also considered some interesting examples which have been
recently discussed in the literature \cite{key-5,key-55,key-56,key-57}.

However, further and accurate studies are needed in order to test
the relation between Sachs-Wolfe effect and f(R) gravity. This goal
could be achieved in the next future through the forthcoming space
(LISA) and ground based (Virgo, Advanced LIGO) interferometers and
Li-Baker HFGW detectors.

\section{Conclusions}

This paper has been a review of previous work on the SBGWs which has
been discussed in various peer-reviewed journals and international
conferences. The SBGWs has been analysed with the aid of the WMAP
data while, in general, in previous works in the literature about
the SBGWs, old COBE data were used. After this, it has been shown
how the SBGWs and f(R) gravity can be related, showing, vice versa,
that a revealed SBGWs could be a powerful probe for a given theory
of gravity. In this way, it has also been shown that the conform treatment
of SBGWs can be used to parametrize in a natural way f(R) theories.
Some interesting examples which have been recently discussed in the
literature, see \cite{key-5,key-55,key-56,key-57}, have been analysed.

The presence and the potential detection of the SBGWs is also quite
important in the framework of the debate on high-frequency gravitational
waves (HFGWs). The importance of HFGWs has been recently emphasized
in some papers in the literature and the ground based Li-Baker HFGW
has been proposed to be sensitive at $10^{10}$ GHz to $10^{-32}$
amplitude HFGWs \cite{key-48}.

\section{Acknowledgments}

I strongly thank the referees for helpful advices which permitted
to substantially improve the paper.

Maria Felicia De Laurentis has to be thanked for good advices too.

This paper has been partially supported by the Sezione Scientifico-Tecnologica
of 0574news.it, via Sante Pisani 46, 59100 Prato, Italy


\begin{thebibliography}{57}
\bibitem{key-1}Peebles PJE and Ratra B - Rev. Mod. Phys. \textbf{75}
8559 (2003)

\bibitem{key-2}E. Elizalde, S. Nojiri, and S.D. Odintsov - Phys.
Rev. D 70, 043539 (2004); G. Cognola, E. Elizalde, S. Nojiri, S.D.
Odintsov and S. Zerbini - J. Cosmol. Astropart. Phys. JCAP0502(2005)010

\bibitem{key-3}G. Allemandi, A. Borowiec, M. Francaviglia, S. D.
Odintsov - Phys. Rev. D 72, 063505 (2005); G. Allemandi, A. Borowiec,
M. Francaviglia, - Phys. Rev. D70 (2004) 103503; 

\bibitem{key-4}E. Elizalde, P. J. Silva - Phys. Rev. D78, 061501
(2008); G. Cognola, E. Elizalde, S. Nojiri, S.D. Odintsov and S. Zerbini
- Phys. Rev. D 73, 084007 (2006)

\bibitem{key-5}G. Cognola, E. Elizalde, S. Nojiri, S.D. Odintsov,
L. Sebastiani, S. Zerbini - Phys. Rev. D 77, 046009 (2008) 

\bibitem{key-6}T.P. Sotiriou and V. Faraoni - arXiv:0805.1726; S.
Capozziello and M. Francaviglia - Gen. Rel. Grav. 40, 2-3, (2008)

\bibitem{key-7}K. Bamba, S. Nojiri and S. D. Odintsov - J. Cosmol.
Astropart. Phys. JCAP10(2008)045

\bibitem{key-8}Will C M \textit{Theory and Experiments in Gravitational
Physics}, Cambridge Univ. Press Cambridge (1993)

\bibitem{key-9}Acernese F et al. (the Virgo Collaboration) - Class.
Quant. Grav. 23 8 S63-S69 (2006) 

\bibitem{key-10}Corda C - Astropart. Phys. \textbf{27,} No 6, 539-549
(2007)

\bibitem{key-11}Hild S (for the LIGO Scientific Collaboration) -
Class. Quant. Grav. \textbf{23} 19 S643-S651 (2006)

\bibitem{key-12}Willke B et al. - Class. Quant. Grav. \textbf{23}
8S207-S214 (2006) 

\bibitem{key-13}Sigg D (for the LIGO Scientific Collaboration) -
www.ligo.org/pdf\_public/P050036.pdf

\bibitem{key-14}Abbott B et al. (the LIGO Scientific Collaboration)
- Phys. Rev. D 72, 042002 (2005) 

\bibitem{key-15}Ando M and the TAMA Collaboration - Class. Quant.
Grav. \textbf{19} 7 1615-1621 (2002)

\bibitem{key-16}Tatsumi D, Tsunesada Y and the TAMA Collaboration
- Class. Quant. Grav. \textbf{21} 5 S451-S456 (2004) 

\bibitem{key-17}Capozziello S and Corda C - Int. J. Mod. Phys. D
\textbf{15} 1119 -1150 (2006); Corda C - \textit{Response of laser
interferometers to scalar gravitational waves}- talk in the \textit{Gravitational
Waves Data Analysis Workshop in the General Relativity Trimester of
the Institut Henri Poincare -} Paris 13-17 November 2006, on the web
in www.luth2.obspm.fr/IHP06/workshops/gwdata/corda.pdf

\bibitem{key-18}Corda C - J. Cosmol. Astropart. Phys. JCAP04009 (2007)

\bibitem{key-19}Corda C - Astropart. Phys. 28, 247-250 (2007)

\bibitem{key-20}Corda C - Mod. Phys. Lett. A 22, 16, 1167-1173 (2007)

\bibitem{key-21}Capozziello S, Corda C and De Laurentis MF - Mod.
Phys. Lett. A 22, 15, 1097-1104 (2007)

\bibitem{key-22}\foreignlanguage{italian}{C. L. Bennet and others
- ApJS \textbf{148} 1 (2003)}

\bibitem{key-23}\foreignlanguage{italian}{D. N. Spergel and others
- ApJS \textbf{148} 195 (2003)}

\bibitem{key-24}G. Garcia-Cuadrado - J. Phys. A 39, 6401-6406 (2006) 

\bibitem{key-25}B. Allen - Proceedings of the Les Houches School
on Astrophysical Sources of Gravitational Waves, eds. Jean-Alain Marck
and Jean-Pierre Lasota (Cambridge University Press, Cambridge, England
1998).

\bibitem{key-26}\foreignlanguage{italian}{L. P. Grishchuk and others
- Phys. Usp. 44 1-51 (2001)}

\bibitem{key-27}\foreignlanguage{italian}{L. P. Grishchuk and others
- Usp. Fiz. Nauk 171 3 (2001)}

\bibitem{key-28}\foreignlanguage{italian}{Babusci D, Foffa F, Losurdo
G, Maggiore M, Mattone G and Sturani R - Virgo DAD - www.virgo.infn.it/Documents/DAD/stochastic
background}

\bibitem{key-29}Corda C - \textit{Virgo Report:} VIRGO-NOTE-PIS 1390-237
(2003) \foreignlanguage{italian}{- www.virgo.infn.it/Documents}

\bibitem{key-30}R. K. Sachs and A. M. Wolfe - \foreignlanguage{italian}{ApJ
147, 73 (1967)}

\bibitem{key-31}\foreignlanguage{italian}{B. Novosyadlyj and S. Apunevych
- proceedings of international confernce {}``Astronomy in Ukraine
- Past, Present, Future'' - Main Astronomical Observatory (2004)}

\bibitem{key-32}J. P. Zibin, D. Scott and M. White \foreignlanguage{italian}{-
arXiv:astro-ph/9904228}

\bibitem{key-33}B. Allen - Phys. Rev. D \foreignlanguage{italian}{\textbf{3}}\textbf{7},
2078 (1988)

\bibitem{key-34}B. Allen and J. P. Romano - Phys. Rev. D \foreignlanguage{italian}{\textbf{59}
102001 (1999)}

\bibitem{key-35}E. E. Flanagan - Phys. Rev. D \foreignlanguage{italian}{\textbf{48}
2389 (1993)}

\bibitem{key-36}K. G. Arun, B. R. Iver, B. S. Sathyaprakash, and
P. A. Sundararajan - Phys. Rev. D \foreignlanguage{italian}{\textbf{71}
084008 (2005)}

\bibitem{key-37}www.lisa.nasa.org; www.lisa-scienze.org 

\bibitem{key-38}S. Weimberg - \textit{Gravitation and Cosmology}
(Wiley 1972)

\bibitem{key-39}Maggiore M and Nicolis A - \foreignlanguage{italian}{Phys.}
Rev. \foreignlanguage{italian}{D \textbf{62} 024004 (2000)} 

\bibitem{key-40}Tobar ME, Suzuki T and Kuroda K \foreignlanguage{italian}{Phys.}
Rev. \foreignlanguage{italian}{D 59 102002 (1999)}

\bibitem{key-41}Corda C - Mod. Phys. Lett. A 22, No. 23, 1727-1735
(2007)

\bibitem{key-42}Capozziello S, Cardone VF, Carloni S and Troisi A
- Int. J. Mod. Phys. D \textbf{12} 1969 (2003)

\bibitem{key-43}Chiba T, Smith TL and Erickcek L - astro-ph/0611867
(2006)

\bibitem{key-44}Starobinsky AA - Phys. Lett. B \textbf{91}, 99 (1980)

\bibitem{key-45}Starobinsky AA - Sov. Phys. JEPT Lett. B \foreignlanguage{italian}{\textbf{34}},
438 (1982)

\bibitem{key-46}Capozziello S, Cardone VF and Troisi A - Phys. Rev.
D \foreignlanguage{italian}{\textbf{71} 08}43503 (2005)

\bibitem{key-47}Capozziello S, Cardone VF , Funaro M and Andreon
S - Phys. Rev. D \foreignlanguage{italian}{\textbf{70} 123501} (2004)

\bibitem[48]{key-48}F. Li, R. M. L. Baker Jr., Z. Fang, G. V. Stephenson
and Z. Chen - Eur. Phys. Journ. C, 56, 3, 407-423 (2008)

\bibitem[49]{key-49}G. Fontana, P. Murad and R.M.L. Baker Jr. - AIP
Conference Proceedings 880, 1117-1124, Melville, New York (2007)

\bibitem[50]{key-50}P. Murad and R.M.L. Baker Jr. - Propulsion Conference
Proceedings 2003-4882, Alabama, USA (2003)

\bibitem[51]{key-51}R. M. L. Baker Jr., G. V. Stephenson and F. Li,
- AIP Conference Proceedings 969, 1045-10544, Melville, New York (2008)

\bibitem[52]{key-52}C. Corda - Astropart. Phys., 30, 209-215 (2008)

\bibitem[53]{key-53}Private communication with the referees

\bibitem[54]{key-54}S. Nojiri and S.D. Odintsov - hep-th 0601213
(2006); S. Nojiri and S.D. Odintsov - ECONFC0602061:06 (2006); S.
Nojiri and S.D. Odintsov - Int. J. Geom. Meth.Mod.Phys.4:115-146 (2007) 

\bibitem[55]{key-55}S. Nojiri and S. D. Odintsov - Phys. Lett. B,
\textbf{657}, 238 (2008), also in arXiv:0707.1941(2007)

\bibitem[56]{key-56}S. Nojiri and S. D. Odintsov - Phys. Rev. D \foreignlanguage{italian}{\textbf{77}
026007} (2008), also in arXiv:0710.1738 (2007)

\bibitem[57]{key-57}S. Capozziello, M. De Laurentis, S. Nojiri, S.
D. Odintsov - arXiv:0808.1335 (2008)
\end{thebibliography}
\end{document}